# EEG for fatigue monitoring

Ildar Rakhmatulin

Miruns, miruns.com,

email: ildarr2016@gmail.com

**Abstract**
Physiological fatigue, a state of reduced cognitive and physical performance resulting from prolonged mental or physical exertion, poses significant challenges in various domains, including healthcare, aviation, transportation, and industrial sectors. As the understanding of fatigue's impact on human performance grows, there is a growing interest in developing effective fatigue monitoring techniques. Among these techniques, electroencephalography (EEG) has emerged as a promising tool for objectively assessing physiological fatigue due to its non-invasiveness, high temporal resolution, and sensitivity to neural activity. This paper aims to provide a comprehensive analysis of the current state of the use of EEG for monitoring physiological fatigue.

## 1. Introduction

Since 1878 the French physiologist Angelo Mosso [52] has carried out pioneering studies of the blood circulation in the brain during mental and physical work, initiating an understanding of the physiological basis of fatigue and the study of physiological fatigue, research efforts have already spanned several disciplines, including psychology, physiology, neurology, and occupational health. Over the years, scientists and researchers have made significant contributions to understanding the nature, causes, and consequences of physiological fatigue. The prediction of physiological fatigue is critical in areas where performance, safety, human well-being and especially sports are of paramount importance. By understanding and predicting fatigue levels it is possibly take proactive steps to reduce fatigue-related risks, optimize performance, and improve overall health and safety.

Today, the prediction of fatigue is required in huge quantities of areas from the prediction of the tiredness of the driver, which is one of the main factors related to traffic accidents [60] and ending with sports implementations. For example, it is known that several players of the Italian football team, who won the 2006 World Cup, used a neurobio management to improve their results [1]. More details about the history and physiology of fatigue, and the feasibility of its detection are described in the work [54, 55]. There are also a large number of survey works in which the authors examined various factors affecting fatigue [57], [58]. Hooda et al. [56] presented this topic more locally and considered the machine learning techniques for detecting fatigue, and Wang et al. [59] did similar work but with an emphasis on the EEG signal.

This paper summarizes the experience of detecting fatigue with an accent to the EEG signals. This paper will help to understand what feature extraction and which technique is advisable to use in the EEG signal to detect fatigue for future research. In recent years, interest in EEG has increased due to machine learning, which greatly simplifies feature extraction from EEG signals [69, 70], and therefore the use of EEG for real-time fatigue prediction is promising with progress in topic of signal processing and feature extraction [] considering advances in hardware [66, 67, 68].

## 2. Short introduction to EEG

It is a non-invasive method used to measure the electrical activity of the brain. Brain cells communicate with each other using electrical impulses, and the EEG records these electrical signals using several electrodes placed on the scalp. During an EEG, the electrodes detect electrical activity generated by neurons in the brain. These electrical signals are amplified and recorded, creating a graphical representation called an electroencephalogram. An EEG recording shows patterns of electrical activity known as brain waves, which can provide valuable information about brain function and activity. For example, delta waves occur in the 0.5 Hz to 4 Hz frequency range and are present during deep sleep, while beta waves occur in the 13 Hz to 30 Hz range and are associated with active thinking. Similarly, other waves are associated with - alpha waves (8-12 Hz): normal waking conditions, gamma waves (30-80 Hz): integration of sensory perception and theta waves (4-7 Hz) [61].

## 3. What is fatigue?

Fatigue can manifest itself in different ways, in this paper we focus on physiological fatigue, but at the same time, we will mentions other types of fatigue since they are correlated with each other:

- **Physical fatigue.** Physical fatigue is the best known type and is often caused by prolonged physical activity or exertion. This can be caused by factors such as muscle fatigue, lack of physical fitness, or overexertion;
- **Mental fatigue.** Mental fatigue is associated with cognitive activity and prolonged mental effort. This may be the result of tasks that require concentration, problem solving, or intense concentration. Mental fatigue can affect cognition, concentration, and decision making. Leshko [8] et al detected Influences of mental fatigue on physical fatigue;
- **Emotional fatigue.** Emotional fatigue is characterized by a feeling of emotional exhaustion and is often associated with prolonged periods of stress, anxiety, or emotional tension. This can be caused by factors such as relationship problems, stress at work, or personal issues. **Emotional fatigue** also has impact on fatigue [9];
- **Chronic Fatigue:** Chronic fatigue refers to persistent and prolonged fatigue that is not relieved by rest or sleep. This is often accompanied by other symptoms such as weakness, memory problems, and difficulty concentrating. Chronic fatigue syndrome (CFS) is a condition that causes extreme and unexplained fatigue and obviously has impact to fatigue [10].

But at the same time, fatigue is affected by a huge number of factors, for example, stress affects fatigue [11] and there are a number of works that demonstrate how to detect stress through EEG [12]. Fatigue is affected by drowsiness [15] which can be easily detected through the EEG [13] as well as insomnia [14]. The patterns can be very different, even in unexpected forms, for example, avoiding caffeine increases the rate of cerebral blood flow and changes the activity of quantitative electroencephalography (EEG), and the rate of cerebral blood flow affects the Middle cerebral artery blood velocity during running [23]. And there is a huge variety of subtypes of not explicitly expressed patterns for Heavy legs, Breathing difficulty, Decreased motivation, Muscle cramps etc.

## 3.Measuring Fatigue, current stage

There are dozens of new devices that can be used to measure athlete fatigue. Persons et al. [7] presented fatigue testing of wearable sensor technologies and considered challenges and opportunities, Fig.1.

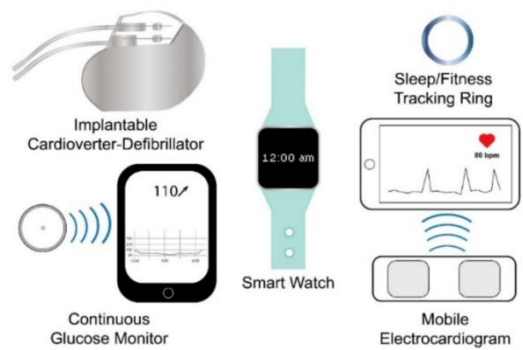

Fig.1. Examples of wearable sensor devices [7]

Budi et al. [4] presented a complete overview of the subject of wearable technology for long-distance runners. The authors concluded that there are currently no standards for fatigue testing of wearable sensors which makes it difficult to determine how useful they are for assessing fatigue.

Next, we will consider the parameters that can be used for fatigue detection:

- **Heart rate monitor**. Methods for measuring heart rate can be based on various technologies such as electrocardiography, photoplethysmography, or accelerometry. Measuring the electrical activity of the heart using an electrocardiogram (ECG) can provide important information about the heartbeat. This indicator is one of the most used for the running process [24] and has a relationship with fatigue [25], [26].
- **PPG (Photoplethysmography)** sensors has been widely used since the introduction of smartwatches [27] and is now widely used by athlels for health monitoring. PPG monitors blood oxygen saturation (SpO2), the percentage of hemoglobin in the blood that is saturated with oxygen. This device can also be used for Heart Rate (HR). A general overview of the capabilities of the PPG sensor is presented in paper [29], these indicators have a correlation with fatigue [28].
- Also like and **Accelerometers** sensors which can be used to predict fatigue in certain contexts by monitoring the movement patterns and activity levels of individuals. While accelerometers alone may not directly measure fatigue, they can provide valuable data that, when combined with other information, can be used as an indicator or predictor of fatigue [62].
- **Seismocardiography** (SCG) refers to the measurement of the vibration of the body and in particular, the chest, caused by the contraction of the heart and the ejection of blood from the ventricles. Currently, it is possible to register SCG by placing a MEMS accelerometer on the chest of a person, this indicator also correlates with fatigue [63] also like and information about **Breathing rate [64]**.
- **Electrodermal activity and sweating (EDA).** Many devices have electrical sensors that can be used for ECG and/or electroskin activity monitoring. EDA, which can also be called galvanic skin response, is a measure of changes in the electrical conductivity of the skin. Sweating is widely used to relieve fatigue [30]. Sweating can be an indicator of physical fatigue, although it is not a direct indicator of fatigue itself. During physical activity or under conditions of stress, the body's thermoregulatory system is activated to maintain a stable body temperature. However, it is important to note that sweating alone is not a definitive predictor of physical fatigue. Other factors such as ambient temperature, humidity, hydration levels, and individual variability can also affect sweating. In addition, there are individual differences in the rate and pattern of sweating that may influence the correlation between sweating and fatigue [33]. Nielsen [38] showed how brain activity and fatigue correlate during prolonged physical exertion in the heat.
- **Glucose.** Continuous blood glucose monitors (CGMs) can provide real-time information about how our bodies respond to generate the energy needed to perform activities such as running. They can be used to assess whether a meal plan is correct or needs to be changed, since each organism reacts differently to incoming food and is correlated with fatigue [31]. Keeping blood glucose levels stable can help prevent fatigue and other issues during the marathon [32]. At the same time, it is obvious that environmental parameters also have a relationship with fatigue, and here everything is individual [34]. As well as anthropometric measurements which represent the dimensions of the human body and skeleton [5].
- The **indicator of fitness** has a direct effect on the fatigue performance. [39]. Relationships between biological, training, and physical variables in the expression of performance in lay runners were considered in the paper [40].

In many papers showed a correlation between biological parameters and fatigue. Fatigue was influenced by physiological parameters such as maximal oxygen uptake (VO 2 max), speed at maximal oxygen uptake, running economy (RE), and changes in lactate levels [6]. Leptin and endocrine parameters in marathon runners [35], Arterial stiffness and wave reflections in marathon runners [36], partial pressure of O2 and CO2 in the air, is the arterial partial pressure of CO2 (PaCO2), which can be estimated by means of the transcutaneous partial pressure of CO2 (PtCO2) [37].

There are hundreds of wearable devices that can measure various indicators, but so far there is no biological parameter by which it would be possible to measure physiological conditions. Each parameter has a relationship between fatigue, but each parameter has an individual parameter and therefore cannot be generalized and used as a standard for determining fatigue.

## 3. Measurement of fatigue in sports, current stage

There are several experimental studies in which the authors tried to apply the knowledge of the work of the brain through the EEG to the effectiveness of the actions of professional athletes. Rijken et al., [41] tried to identify patterns between the performance improvement of professional football players and elite track and field athletes with the help of maximum performance training and biofeedback that changes in EEG parameters. The authors found that as a result, significant changes over time in the power of the alpha channel in the EEG. A similar result was obtained in a study investigating the effect of alpha power training on gymnasts, small ones were found where non-significant changes in alpha signal strength were found after training in athletes [2]. Chuang et al., [65] discovered a useful correlation in the theta range. Fluctuations in low and high theta power values have been found to be a predictor of shot misses in basketball players, suggesting that sustained attention and arousal may contribute to athletic performance.

In a study [42], the authors assessed the impact of holistic training on the physiological (EEG) and behavioral performance of semi-professional athletes. The influence of training showed an increase in the amplitudes of SMR (12-15 Hz) and beta1 (13-20 Hz) ranges and a simultaneous decrease in the amplitude of theta (4-7.5 Hz) and beta2 (20-30 Hz).

Pineda-Hernández [43] found the correlation between fatigue and gamma waves. In this study consisted of observing activation during neutral situation imagery (NSI) and pressure situation imagery (PSI) based on the analysis of heart rate, brain waves, and subjective assessments in athletes. In PSI, an increase in gamma waves was also observed in the interval 5–6, the moment of maximum pressure. Bieru et al.,[44] already found a correlation in theta waves. In this work, authors studied brain activity using electroencephalography (EEG) and assessed differences in brain functions depending on specific sports activities. The authors conducted a study on two sets of athletes in judo (12 subjects) and volleyball (11 subjects). As a result, the values of theta waves are slightly increased in judokas, and we also observe a correlation between alpha waves for the dominant hemisphere. Alvero-Cruz [7] opposite used not the traditional frequency for EEG signals. The authors considered the EEG activity and mood of health-oriented runners after exercises of different intensities. The results showed the effect of preferred speed and high-intensity speed on both EEG and mood. Since only areas with higher N18 Hz frequency showed a persistent decrease after training, authors concluded that this may be a sign of a long-term effect of exercise on cortical activity.

As can be seen from the results, not only is there no single or recognized standard for detecting fatigue through EEG readings, but there is not even a complete understanding of what frequency to look for a correlation between EEG data and fatigue. But for the most part, the authors come to the conclusion that changes in signal spectral power in the alpha range are a reliable indicator of subject fatigue.

## 4. Measurement of fatigue for running, current stage

Running and EEG is a very popular topic, since in addition to running, a huge number of people go in for running for their health. In the EU alone, about 45-55 million people are involved in running activities, Bridveld [3]. and therefore several interesting papers have been found on this topic, in which the fatigue was detected by EEG in runners. The results of the study showed that in well-trained and acclimatized athletes, arousal in EEG signals plays a protective role in preventing excessive oxygen starvation even after endurance exercises performed at high altitudes [21].

In the paper [45] twelve runners wore portable EEG headsets throughout the race, and post hoc analysis showed changes in mental state related to distance, finish time, gender, age, experience, climb, and recovery loads. The main components of the regression have been confirmed by retrospective studies stating that the marathon consists of three psychological stages. Runners are more alert in the first leg, then begin to transition to the rhythm at mile 11 and show an internal mental shift from mile 18 to the finish line. Men showed higher frontal theta (i.e. internal attention) in the second stage, while women showed higher global alpha (i.e. relaxation).

Moussiopoulou et al., [46] studied the effects of excessive exercise during marathon running. The study included 30 healthy marathon runners. EEG channel recording was carried out: 12–8 weeks before the marathon run, 14–4 days before the marathon run, 1–6 days after, and 13–15 weeks after the marathon run. Statistical non-parametric mapping showed a decrease in power in both the alpha and delta frequency bands after the marathon run.

In work [48], EEG data were taken from marathon runners during a two-week endurance training (i.e., at the beginning, after one and after two weeks) and processed by the off-line method of fast spectral analysis. EEG showed in both groups a decrease in theta activity (4-6 Hz) and an increase in the slow alpha component and subtheta activity (6-8 Hz) without fundamental differences between the experimental and control groups. The changes may be related to fatigue and possibly poorer oxygen and glucose supply.

Sauseng et al., [49] recorded EEGs data during 12-, 24- and 56-hour ultramarathons. Analysis of the frequency of the center of gravity of the EEG alpha rhythm showed a gradual decrease with time for 12- and 24-hour races. In the 56-hour race, the central frequency decreased only until the first sleep period. The alpha amplitude, on the other hand, did not change systematically. For all three races, the lowest alpha amplitude was observed during the last test session.

Aaiswal et al.,[18], created an experimental setup that included multiple standardized cognitive tasks and a physical task (running on a treadmill). As participants completed each of the assigned tasks, data modalities such as ECG (for heart rate variability), EMG (for muscle activation), EDA/GSR (for skin conduction and emotional arousal), and EEG (for electrical brain signals) were recorded. Self-assessment scores were analyzed to check for fatigue in the participants. Finally, several statistical features were extracted from the collected signals and machine learning (ML) models were trained to predict participants' fatigue levels. While the Random Forest classifier performed best in PF detection, the success of the LSTM model in CF prediction eliminates the need for extensive data preprocessing and feature extraction, Fig.2.

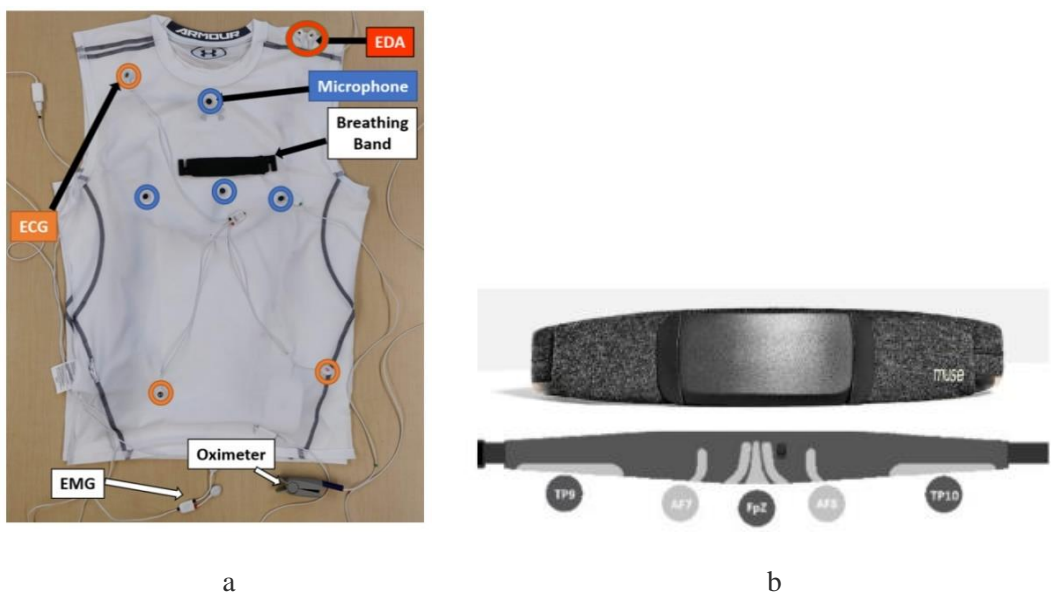

a        b

Fig.2. a - a prototype of the sensor shirt with different type so sensors, b – muse device for EEG

This work is an excellent example of collecting a dataset to identify correlations between fatigue and body performance. Only in this work, the data was used for machine learning, and although a positive result was obtained, no fundamental dependencies were examined and no clearer correlations were found between biological data in the processes of determining fatigue.

**5. Fatigue prediction for drivers, current stage**
Fatigue is one of the main causes of road traffic accidents, therefore, the topic of detecting fatigue in this area using EEG signals is a very promising direction [17].
Chen [16], proposed an algorithm for detecting mental fatigue based on the extraction and fusion of multidomain features. The EEG components representing fatigue are closely linked past and present because fatigue is a dynamic and gradual process. Accordingly, the idea of linear prediction is used to compare the current value with a set of sample values in the past to calculate cepstral linear prediction coefficients (LPCC) as a feature of the time domain Cepstral linear prediction coefficients for EEG (electroencephalogram) signals are a set of features that capture the spectral characteristics of the signal EEG using linear prediction analysis. Cepstral coefficients provide a representation of the logarithm of the power spectrum of a signal and can be used for various analysis tasks such as feature extraction, classification or anomaly detection, and the detection of static in EEG data [51].
In the paper [19], in order to study the state of fatigue when driving under conditions of low voltage and hypoxic plateau, the study included aspects of subjective monitoring and objective monitoring according to the electroencephalogram (EEG) signal of the driver in real time, which was obtained during the field test for fatigue during driving. Nonlinear and linear methods were used to analyze the EEG signal in three typical states of wakefulness, critical state and fatigue. EEGLAB in the MATLAB toolbox was used in a non-linear manner to analyze the power spectral density map of θ, α, β waves in three typical states. New EEG eigenvalues were collected and EEG power characteristic values were calculated to estimate the trend of the EEG signal by a linear method. Combined non-linear and linear methods with analysis of subjective data, the energy response (α+θ)/β and (α+β)/θ have been recommended as an indicator for assessing fatigue performance in low stress and hypoxic plateau driving conditions. The authors of this work presented the EEG spectrum and topographic map for the brain of three states, Fig. 3.

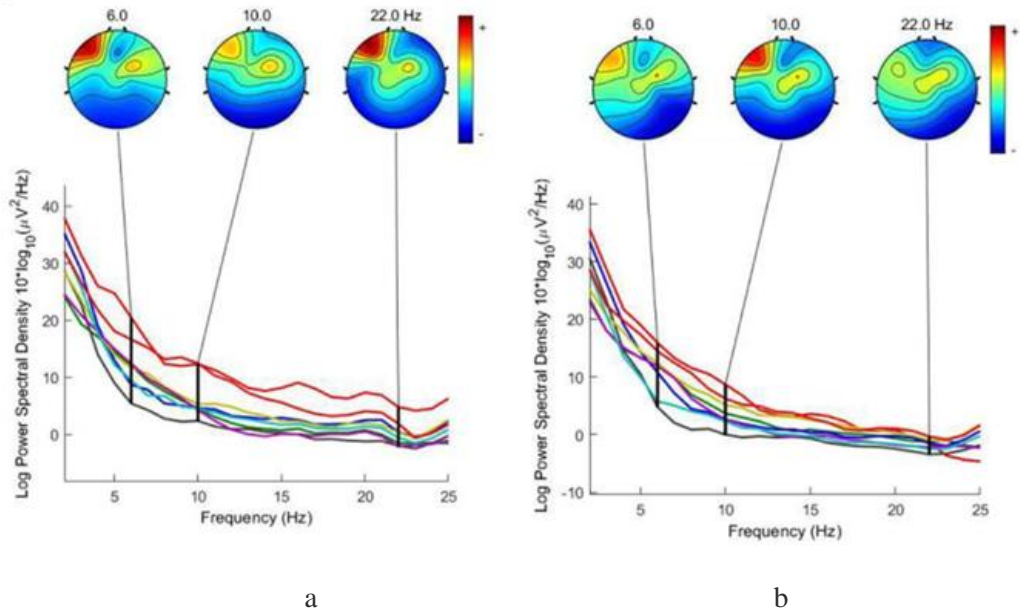

a  b
Fig. 3. EEG spectrum and brain topographic map of three states: a - awake state, b - fatigue state

The spectrograms show that in the process of transition from the state of wakefulness to the state of fatigue, the amplitude tends to smoothness. The amplitudes of the α- and β-waves decrease, while the amplitude of the θ-wave has no obvious changes.
Tey et al., [20], presented the first study a 4-channel dry EEG in a driving simulation to predict when a driver will develop microsleep over the next 10 minutes using only 3-minute data. The authors focused on spindle detection. Spindle EEG refers to a specific pattern of brain activity that is observed when an electroencephalogram (EEG) is recorded. spindle is characterized by short bursts of oscillatory brain waves in the frequency range of 8 to 16 hertz (Hz), typically 0.5 to 2 seconds long. In this paper, spindles appear during early sleep states in the 9-15 Hz range of EEG signals. Thus, a simple measurement of the spindles can provide a level of fatigue at a given time. The spindle detection algorithm achieved an accuracy of 91.71% with traditional wet EEG data and a lower accuracy of 89.31% was achieved with dry EEG sensors.

Similar research was presented by Gritti et al. [21] but is more intriguing because it uses an EEG and gyroscope sensor. Finally, the authors proposed a hybrid multimodal system based on a neural network that determines driver fatigue using electroencephalography (EEG) data, gyroscope data, and image processing data from the camera. The proposed hybrid system was found to perform well with a detection accuracy of 93.91% in detecting driver drowsiness.

Budi et al. [53] examined various physiological associations with fatigue in an attempt to identify indicators of driver fatigue. The results showed a stable delta and theta activity over time, a slight decrease in alpha activity, and a significant decrease in beta activity ($p<0.05$).

Research on predicting fatigue is more focused on practical results, the resulting dataset is used for machine learning algorithms. The resulting models are eventually able to predict fatigue in the control group, but there is no complete understanding of how this model will work when data recording conditions change and how fatigue is detected, according to what patterns. Therefore, a much more important point is the fundamental understanding of the fatigue process through EEG data in order to subsequently detect a certain parameter for predicting fatigue.

**Conclusion**

Summarizing, we can say that there are three directions in the processes of fatigue detection, one of the very first formed is the study of various frequency ranges for finding and comparing the activity of the EEG signal before and after the onset of fatigue. A number of studies have shown that alpha activity is the most sensitive indicator that can be used to detect fatigue, followed by theta and delta activity [4, 47].

Then using paradigms such as the spindle, these studies l ook very promising as they suggest detecting a biological pattern that is generally stable when measuring the EEG signal in different subjects. But there is not enough research in this area. The last case is the use of machine learning. Artificial intelligence requires a large amount of dataset, but even a positive result does not give an understanding of how the function is extracted. Machine learning can detect fatigue via EEG signals, but there is a "black box" problem of not understanding how the model works and how it will perform outside of the control group.

Given the greater number of additional wearable biosensors, the logical direction in this study is to work on collecting datasets from biosensors and EEG data before and after physical exercises. It is necessary to work on the structuring of the study, the creation of a map of the variation between the EEG and other indicators of fatigue.

The available datasets do not represent all the data about the process of recording a signal from a subject, in fact, an additional dataset should be available to the main dataset with a complete bio-passport of the subject, environmental parameters at the time of reading the data and any other available information.

Although EEG is not usually used directly to monitor fatigue for applied purposes, EEG can nevertheless provide valuable information about brain activity and cognitive states associated with fatigue. However, the EEG has a number of advantages as it provides an objective measure of brain activity, allowing researchers and practitioners to assess the cognitive and neurological aspects of fatigue. It can reveal patterns of brain wave activity associated with various cognitive states such as alertness, drowsiness, and mental fatigue. An EEG can detect changes in brain wave patterns that occur during states of fatigue. By monitoring these changes, the EEG can display fatigue levels in real-time, allowing athletes and coaches to make adjustments to training strategies or pace. EEG can be used for biofeedback training, which involves giving people real-time feedback about their brain activity. By monitoring their own EEG signals, runners can learn to recognize and modulate their mental state, potentially reducing fatigue. EEG data combined with other physiological and performance data can help create customized training plans for marathon runners. By understanding how an athlete's brain activity changes during different training or race conditions, coaches can optimize training regimes, recovery strategies, and even race pace to minimize fatigue and increase performance. At a minimum, EEG should be considered as part of a comprehensive approach to fatigue management, unless it can be used as a standalone solution.

**Conflicts of Interest**: None
**Funding**: None
**Ethical Approval**: Not required